\documentclass[12pt,letterpaper]{article}
\usepackage[top=1in,bottom=1in,left=1in,right=1in]{geometry}
\usepackage[utf8]{inputenc}
\usepackage[breaklinks,colorlinks,urlcolor=blue,citecolor=blue,linkcolor=blue]{hyperref}
\usepackage{graphicx}
\usepackage{amsmath}
\usepackage{amssymb}
\usepackage{natbib}
\usepackage{multicol}
\usepackage{setspace}

\graphicspath{{figures/}{assets/}}

\newcommand{\incfig}[2][]{%
  \IfFileExists{figures/\detokenize{#2}}{%
    \includegraphics[#1]{\detokenize{#2}}%
  }{%
    \IfFileExists{assets/\detokenize{#2}}{%
      \includegraphics[#1]{\detokenize{#2}}%
  }{%
    \IfFileExists{\detokenize{#2}}{%
      \includegraphics[#1]{\detokenize{#2}}%
    }{%
      \parbox[c][0.18\textheight][c]{\linewidth}{\centering\small\texttt{figures/\detokenize{#2}}\\[0.5em]missing (expected under \texttt{figures/})}%
    }%
  }%
  }%
}

\begin{document}

\pagenumbering{gobble}

\Huge
\begin{center}
Hubble as a Unique Discovery Engine of the Fate of Massive Stars and\\ Black Hole Formation 
\end{center}
\normalsize

\begin{center}
Avishai~Gilkis$^{1}$, Eva~Laplace$^{2, 3, 4}$, Charles~D.~Kilpatrick$^{5}$,\\ Maria~R.~Drout$^{6, 7}$, Anna~J.~G.~O'Grady$^{8}$ and Christopher~A.~Tout$^{1}$\\
\smallskip
\begin{tiny}
$^{1}$ Institute of Astronomy, University of Cambridge, Madingley Road, Cambridge CB3 0HA, United Kingdom\\
$^{2}$ Institute of Astronomy, KU Leuven, Celestijnenlaan 200D, 3001 Leuven, Belgium\\
$^{3}$ Leuven Gravity Institute, KU Leuven, Celestijnenlaan 200D, box 2415 3001 Leuven, Belgium\\
$^{4}$ Anton Pannekoek Institute for Astronomy, University of Amsterdam, Science Park 904, 1098 XH Amsterdam, the Netherlands\\
$^{5}$ Center for Interdisciplinary Exploration and Research in Astrophysics (CIERA), Northwestern University, Evanston, IL 60201, USA\\
$^{6}$ David A. Dunlap Department of Astronomy and Astrophysics, University of Toronto, Toronto, M5S 3H4, ON, Canada\\
$^{7}$ Observatories of the Carnegie Institution for Science, Pasadena, 91101, CA, USA\\
$^{8}$ McWilliams Center for Cosmology and Astrophysics, Department of Physics, Carnegie Mellon University, Pittsburgh, PA 15213, USA\\
\end{tiny}
\end{center}

\bigskip

\noindent
{\bf Abstract:}\\
How stellar-mass black holes are formed is an open question in astrophysics, with very limited observational constraints. It is not known which types of stars are more likely to produce black holes, and whether the formation process is accompanied by strong or weak electromagnetic transients -- or none at all -- and this issue remains a critical missing piece in the puzzle of the fate of massive stars.
Recent theoretical work predicts that many stellar-mass black holes form from hot, UV-luminous massive stars, including Wolf–Rayet-like progenitors, and searches focused primarily on luminous cool supergiants may therefore miss a substantial fraction of black-hole formation events.
While the coming decade will bring major advances in time-domain astronomy through Rubin/LSST, \textit{Roman}, \textit{JWST}, and wide-field transient surveys, none of these combines UV sensitivity, sub-arcsecond imaging, and decade-long continuity. \textit{HST} uniquely enables direct searches for disappearing hot massive stars associated with black-hole formation.
We outline a roadmap for extending \textit{HST}’s role in this area into the 2030s through a dedicated, large program to re-image nearby galaxies in the UV and identify candidate disappearing stars and unusual low-luminosity transients identified by complementary surveys. Theoretical event rates imply that the nearby galaxy population accessible to \textit{HST} should yield of order one detectable black-hole-forming disappearance event per year. Extending \textit{HST} operations into the 2030s would therefore provide crucial insights into the unsolved problem of black hole formation.

\pagenumbering{arabic}

\section{The importance of {\it HST} for imaging UV-luminous BH progenitors}\label{sec:blueness}

Massive stars are the progenitors of supernovae, gamma-ray bursts, neutron stars, black holes (BHs), and many of the compact-object binaries now observed through gravitational waves. The Hubble Space Telescope (\textit{HST}) has been the main instrument establishing the connection between core-collapse supernovae and massive stars through direct imaging of about two dozen identified progenitor stars \citep[e.g.][]{Smartt15,VanDyk2017,VanDyk2023}. While \textit{JWST} will provide optical and infrared detection of supernova progenitors \citep[e.g.][]{Kilpatrick2025}, the crucial UV capabilities of \textit{HST} remain essential for the analysis of the hottest stellar endpoints.

Hundreds of merging stellar-mass BHs have been detected via gravitational waves \citep{LIGO2025O4}, and BHs in binary systems have also been observed for many years \citep[e.g.][]{Remillard2006,Shenar2022,ElBadry2023GaiaBH1}, yet the pathways leading to stellar-mass BH formation \citep[e.g.][]{Burrows2025} remain poorly constrained. In particular, some massive stars may collapse directly to BHs without producing a successful supernova explosion \citep{Laplace2025}, instead presenting observationally as disappearing stars \citep{Kochanek2008,Adams17,De2026}.

Searches for such failed supernovae over the past two decades have largely focused on luminous cool supergiants monitored in optical bands \citep[e.g.][]{Jencson2022}. However, stellar theory strongly suggests that a large fraction of BH progenitors would be hot, blue, and UV-luminous stars, frequently in stripped or Wolf–Rayet-like evolutionary phases \citep{Gilkis2025,Gilkis26BH}.
These progenitors are intrinsically faint \citep[e.g.][]{Yoon2012,Eldridge2013,Drout2023,Gotberg2023} in the red optical bands used by many current surveys and are expected to reside preferentially in crowded star-forming environments, implying that existing searches may be systematically incomplete with respect to the dominant channels of stellar-mass BH formation.

The coming decade will transform nearby time-domain astronomy through facilities including Rubin/LSST, \textit{Roman}, \textit{JWST}, and wide-field optical and infrared transient surveys. These programmes will discover unprecedented numbers of nearby supernovae, stellar eruptions, and disappearing-star events. However, none of these facilities combines deep UV sensitivity with the sub-arcsecond angular resolution required to isolate hot massive stars in crowded nearby galaxies. \textit{HST} uniquely occupies this observational regime through its combination of UV coverage, stable high-resolution imaging, and decades-long archival baseline.

For these reasons we propose that direct searches for BH formation from hot massive stars should become a key component of \textit{HST}’s scientific roadmap into the 2030s.

\begin{figure}[!t]
    \centering
    \incfig[width=0.92\textwidth]{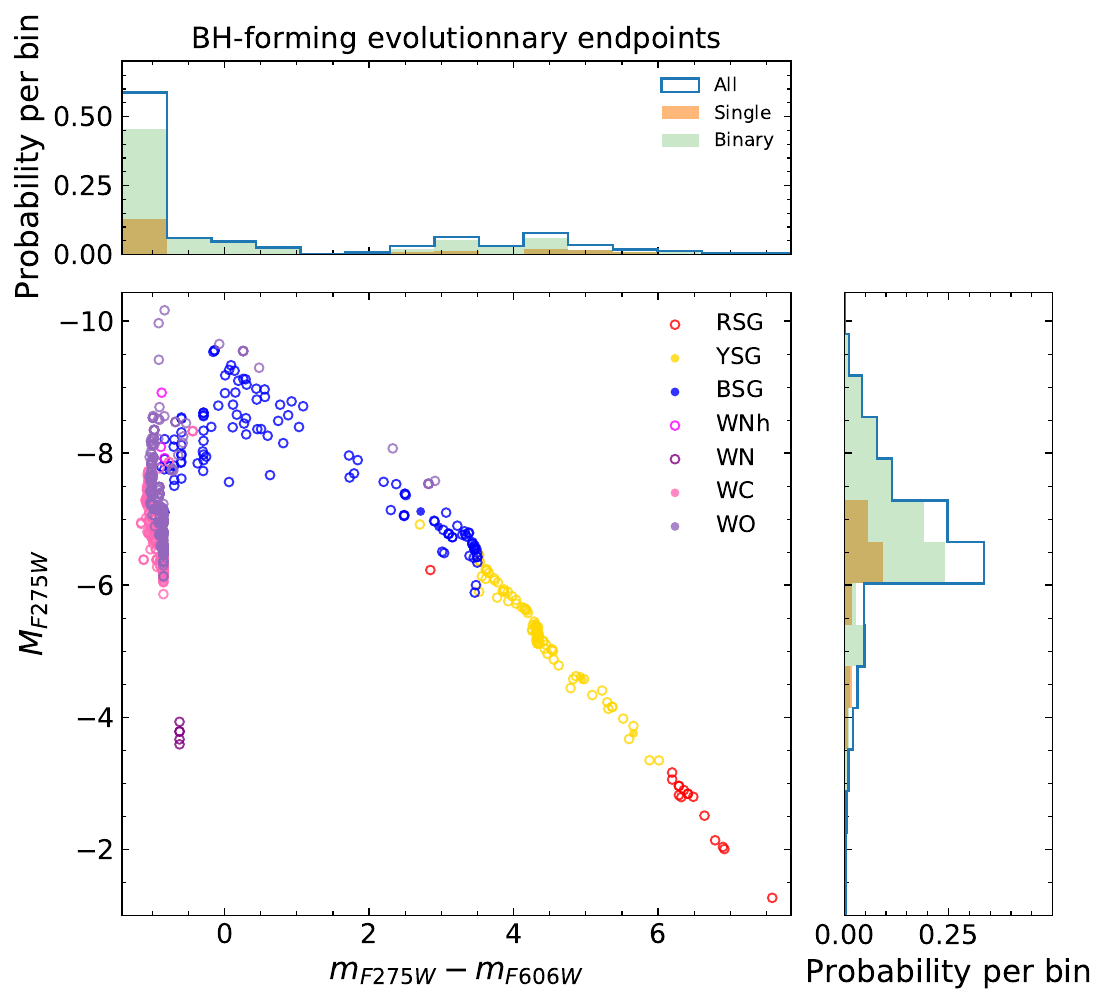}
    \vspace{-8pt}
    \caption{Colour--magnitude diagrams for BH-forming progenitors at Milky-Way-like metallicity, including single star and binary system  endpoints. The plotted photometry includes the flux contribution of any non-degenerate companion present at collapse, while the colour coding/classification of each point is based on the properties of the BH-forming progenitor itself. The absolute magnitude histograms on the side and colour histograms above show weighted distributions, accounting for the initial mass function, assuming a $70\%$ binary fraction, and following empirical distributions of initial orbital configurations.}
    \label{fig:CMD}
\end{figure}

\section{Why {\it HST} is uniquely required}\label{sec:uniqueness}

Only \textit{HST} combines the UV sensitivity, high angular resolution, and long-term temporal continuity needed to find the majority of BH-forming massive stars in the nearby Universe. No existing or planned facility provides this capability combination -- Rubin/LSST, \textit{Roman}, and \textit{JWST} occupy complementary observational parameter spaces. Rubin/LSST will provide an unprecedented transient discovery rate over wide fields, but its ground-based resolution blends stellar populations on scales of tens of parsecs at distances of $\sim10$–$20\,\mathrm{Mpc}$. \textit{Roman} and \textit{JWST} will provide exceptional infrared sensitivity and resolution, but lack access to the UV wavelength range where hot massive stars emit most strongly. Future UV survey missions will provide large-area monitoring capabilities, but without the sub-arcsecond imaging resolution required to isolate individual massive stars in crowded nearby galaxies.

\textit{HST} is therefore exceptionally well suited to UV studies of disappearing massive stars, capable of deep UV imaging with stable $\sim0.1''$ resolution over decade-long baselines. This capability is essential for resolving crowded star-forming environments \citep[e.g.][]{Calzetti15,lee22}, separating massive stars from nearby companions and clusters, and performing precise image subtraction between epochs to identify stars that have vanished or dramatically faded. \textit{HST} complements optical and infrared transient surveys, providing UV identification and characterization of candidate disappearing stars discovered by Rubin/LSST or other facilities, while archival \textit{HST} imaging enables direct comparison against pre-existing UV observations of nearby galaxies.

\begin{figure}[!t]
    \centering
    \incfig[trim={0 10cm 0 0},clip,width=0.92\textwidth]{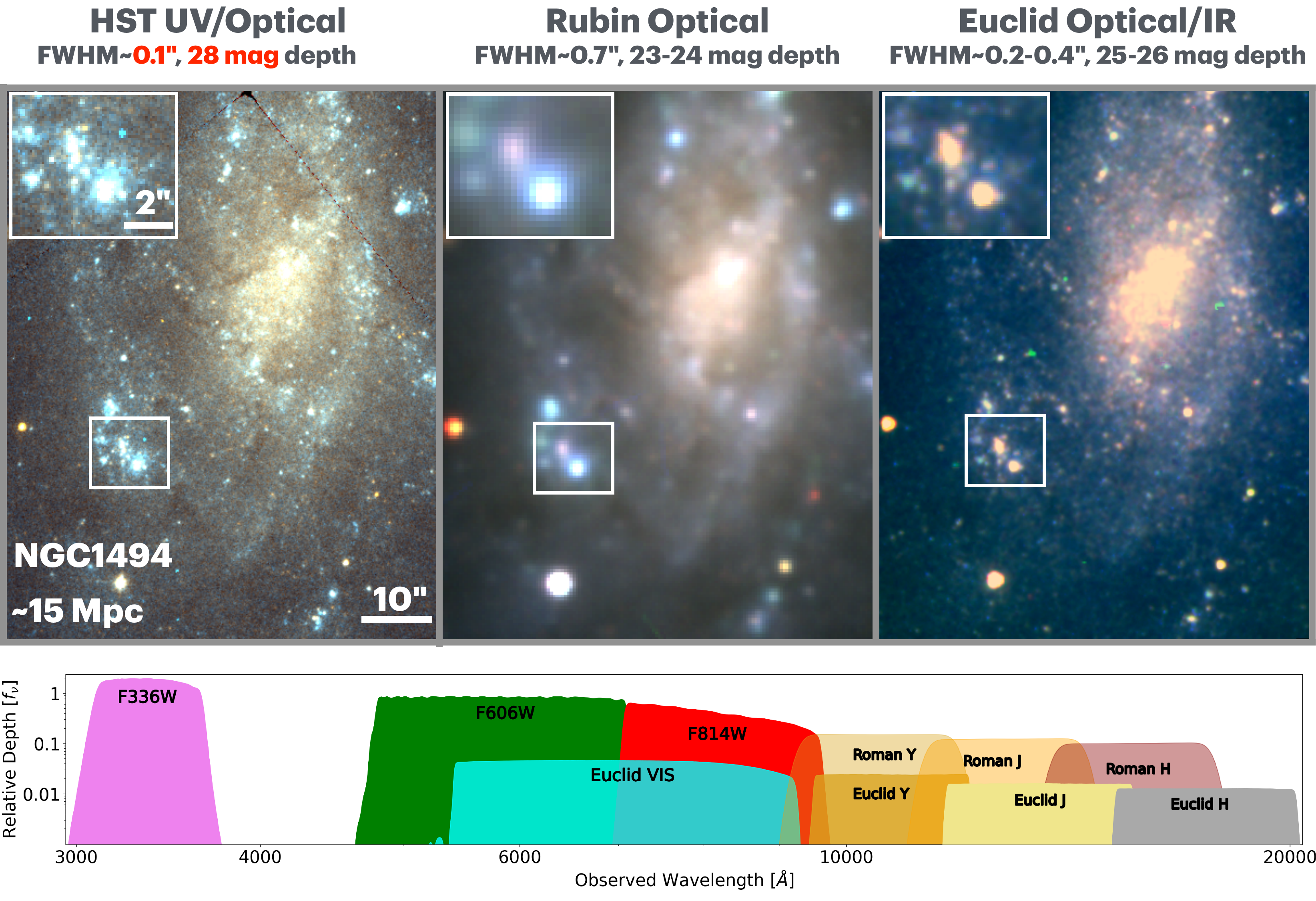}
    \vspace{-8pt}
    \caption{
        Comparison of {\it HST} (F450W+F814W), Rubin (simulated from DECam at similar single-epoch depth), and {\it Euclid} (VIS+$YH$) for NGC\,1494 at $15\,\mathrm{Mpc}$. Superior resolution, depth, and UV/optical coverage enable {\it HST} to detect and deblend more stars than the other facilities (inset). {\it Roman} will have comparable resolution to {\it HST}\ in the IR, but not at UV wavelengths.    
    }
    \label{fig:OBS}
\end{figure}

\section{A Roadmap for Detecting BH Formation with {\it HST}}\label{sec:strategy}

Recent theoretical modelling combining detailed stellar and binary evolution calculations with physically motivated prescriptions for core-collapse outcomes predicts that BH formation through failed or weak supernovae should occur at a non-negligible rate in nearby star-forming galaxies -- using synthetic populations weighted by the initial mass function and empirical binary distributions, predicted direct-collapse rates are approximately $0.4$ events per century for a galaxy forming stars at $1\, M_\odot\,\mathrm{yr}^{-1}$ \citep{Gilkis26BH}. The nearby galaxy population accessible to \textit{HST} therefore implies an expected event rate of order one detectable BH-forming disappearance per year when considering the combined star-formation rate within distances of $\sim10$–$20\,\mathrm{Mpc}$. Importantly, many progenitors are expected to be UV-luminous and located in crowded star-forming environments, making high-resolution UV imaging essential for their detection and interpretation.

\textit{HST} is uniquely suited to exploit this opportunity through two complementary observing modes. First, archival re-imaging of nearby galaxies with existing \textit{HST} UV observations would enable direct searches for stars that have vanished or undergone significant UV fading over decade-long baselines.
Secondly, \textit{HST} provides the UV high-resolution follow-up capability required to characterize candidate disappearing stars and unusual low-luminosity transients identified by Rubin/LSST and other wide-field surveys.

This strategy is particularly timely because extending \textit{HST} operations into the 2030s substantially increases both the available temporal baseline and the expected number of detectable events. Over decade-long timescales, \textit{HST} observations could be the key to answering a major unsolved problem in astrophysics -- the formation of BHs.

\section{Summary}

\textit{HST} uniquely combines UV sensitivity, high angular resolution, and long-term continuity, providing the essential capability for identifying stellar-mass BH formation. Recent theoretical predictions suggest that the nearby galaxy population accessible to \textit{HST} should yield of order one detectable black-hole-forming disappearance event per year in the 2030s. Extending \textit{HST} operations would therefore enable statistically meaningful constraints on black-hole formation channels while complementing Rubin/LSST, \textit{Roman}, \textit{JWST}, future UV missions, and the Habitable Worlds Observatory.

\section*{Acknowledgments}

We thank Andreas Sander for helpful discussions on the unique role of \textit{HST} in studies of hot massive stars and for encouraging the development of this white paper.

\bibliographystyle{yahapj}
\renewcommand{\bibsection}{}
\begin{multicols}{2}[\section*{References}]
\setlength{\bibsep}{0pt plus 0.2pt}
\setstretch{0.88}
\small
\bibliography{references}
\end{multicols}

\end{document}